\documentclass[final,5p,times,twocolumn,preprint]{elsarticle}
\usepackage{graphicx}
\usepackage{slashed}
\usepackage{amsmath,amsfonts,bm,bbold}
\usepackage{amssymb}
\usepackage{calligra}
\usepackage[T1]{fontenc}

\begin{document}

\author{\centerline{A.~V.~Radyushkin}
       }
\address{\centerline{Physics Department, Old Dominion University, Norfolk,
             VA 23529, USA}}
\address{\centerline{Thomas Jefferson National Accelerator Facility,
              Newport News, VA 23606, USA}
}

\title{ \flushright{\large  JLAB-THY-14-1873} \\ 
\begin{center}
Virtuality Distributions in   Application to       $\gamma\gamma^* \to \pi^0$ 
Transition \\ Form Factor at Handbag Level 
\end{center} }

\begin{abstract}

We  outline basics of a  
 new approach to transverse momentum dependence 
 in hard processes.   As an  illustration, we consider 
  hard exclusive transition process
 $\gamma^* \gamma \to \pi^0$  at the handbag level.
 Our  starting point   is  
coordinate representation for  matrix elements
 of operators  (in the simplest case, bilocal  ${\cal O} (0,z)$) 
 describing   a hadron with momentum $p$.
 Treated as  functions of $(pz)$ and $z^2$, they are  parametrized 
 through     {\it virtuality distribution  amplitudes} (VDA)   $\Phi (x, \sigma)$,  with  
 $x$ being  Fourier-conjugate to $(pz)$ and  $\sigma$  
  Laplace-conjugate  to $z^2$.  
 For   intervals with   $z^+=0$,  
 we   introduce the  {\it transverse momentum distribution amplitude}  (TMDA) 
 $\Psi (x, k_\perp)$,   and  write it 
 in terms of VDA   $\Phi (x, \sigma)$. 
The  results of covariant calculations,  
written in terms  of $\Phi (x, \sigma)$ are  converted   into expressions 
involving  $\Psi (x, k_\perp)$. 
   Starting with scalar toy models, we  extend the analysis  
   onto the case of spin-1/2 quarks  and QCD. 
 We propose  simple models for soft VDAs/TMDAs,
and  use them  for comparison of handbag results with experimental 
(BaBar and BELLE)  data 
on  the pion transition form factor. We also discuss  how one can generate 
high-$k_\perp$  tails from primordial soft distributions.


\end{abstract}

\maketitle

\section{Introduction}

Analysis  of effects due to  parton  transverse momentum  is 
an important direction in  modern  studies of hadronic structure. 
The main effort is to use  the transverse-momentum
dependence of inclusive processes, such as 
semi-inclusive deep inelastic scattering (SIDIS) and Drell-Yan 
pair production, describing their cross sections 
in terms of  {\it transverse momentum dependent  distributions}  
   (TMDs)    $f(x,k_\perp)$ \cite{Mulders:1995dh}, which are generalizations
   of the usual 1-dimensional parton densities $f(x)$.
The latter  describe the distribution 
in the fraction $x$ of the longitudinal hadron momentum 
   carried by a parton. 
   
    Within the  {\it operator product
   expansion}  approach (OPE) $f(x)$ is defined \cite{Georgi:1976ve,Efremov:1976ih} 
   as a function 
   whose $x^n$ moments are proportional to matrix elements 
   of twist-2 operators
   containing $n$ derivatives $(D^+)^n$ 
   in the ``longitudinal plus''  direction.
  Analogously, the $(k_\perp^2)^{l}$   moments of TMDs
  correspond to matrix elements of operators
  containing the derivative $(D_\perp^2)^l$  in the transverse direction.
   However, in a usual twist decomposition 
   of  an original  bilocal operator  $\bar \psi (0) \ldots \psi (z)$
   one deals with Lorentz invariant traces of 
   $\bar \psi (0) D_\mu \ldots D^\mu \psi (0)$ type  that correspond 
   to parton distributions in virtuality $k^2$ rather than transverse 
   momentum $k_\perp^2$. Since $D_\perp^2$ is a part of $D^2$,
   it is natural to expect that distributions in transverse
    momentum are 
   related to distributions in virtuality.

 Our goal  is to investigate the relationship 
   between  distributions in virtuality and  distributions in
    transverse  momentum.  We find it simpler to start the study with 
      {\it  exclusive} processes. This  allows to avoid 
    complications specific to  inclusive processes
 (like unitarity cuts, fragmentation functions, etc). 
 Also, among hard  exclusive reactions, we choose the  
  simplest    process  of  $\gamma^* \gamma \to \pi^0$ transition  that 
   involves   just  one hadron. 
Furthermore, 
 this process was studied both 
 in the light-front formalism  \cite{Lepage:1980fj} and in the covariant 
 OPE approach  \cite{delAguila:1981nk,Braaten:1982yp,Kadantseva:1985kb,Musatov:1997pu}.

In our  OPE-type  analysis of the $\gamma^* \gamma \to \pi^0$
process performed in the present paper, we encounter   a TMD-like 
 object, the 
{\it transverse momentum dependent distribution amplitude}
(TMDA) $\Psi (x, k_\perp)$   that is a 3-dimensional 
generalization of the pion  {\it distribution amplitude}  
$\varphi_\pi (x)$ \cite{Radyushkin:1977gp,Efremov:1979qk,Chernyak:1977fk,Lepage:1979zb}. 
The   definition of TMDA $\Psi (x, k_\perp)$ is similar   to 
that 
of TMDs $f(x,k_\perp)$, 
and also to that   of 
 the   pion  wave function used 
in the standard  light-front formalism   
(see, e.g. \cite{Lepage:1980fj}).

We  start in  the next section with an analysis  of 
a scalar handbag diagram. 
We discuss the structure of the  relevant bilocal matrix element 
$ \langle  p | \phi (0)  \phi(z)  | 0 \rangle $  as a function of Lorentz 
invariants $(pz)$ and $z^2$ 
and introduce 
the {\it virtuality distribution amplitude} (VDA)   $\Phi (x, \sigma )$,
the basic object  of our approach. It  describes
the distribution of quarks in the pion both in the longitudinal
momentum 
(the variable $x$ is conjugate
to $(pz)$)  and in  virtuality (the variable $\sigma$ is conjugate
to $z^2$).

 The main features  of our  approach remain intact 
for the  case of spinor quarks, and they also  hold  when the gluons
are treated as gauge  particles (Abelian or non-Abelian). 
For these  reasons, we introduce the basic elements of the VDA  approach
using the  simplest scalar example. In particular,
we show that
the  covariantly   defined
VDA $\Phi (x, \sigma )$
has a   simple connection  to 
 the {\it impact parameter distribution amplitude}  $\varphi (x, z_\perp)$ 
 (IDA) defined for a spacelike interval $z=\{z^-,z_\perp \}$. Then we define 
  TMDA  $\Psi (x, k_\perp)$ as a Fourier transform of 
  IDA $\varphi (x, z_\perp)$.

In Sect. \ref{Spin}, we consider  general  modifications 
that appear for  spin-1/2 quarks, and then  
 show that  the structure of the results 
does not change if one switches further to gauge theories.   
Using  the parametrization 
in terms of VDA 
$\Phi (x, \sigma )$, we calculate the handbag
 diagram  and express the result in terms of the TMDA 
 $\Psi (x, k_\perp)$. 
 In Sect. \ref{TMDA_Models},  we formulate a few simple models for 
soft \mbox{TMDAs,}
and in Sect. \ref{FF_Model} we analyze the application of these models to the 
pion transition form factor.  
In QCD, the quark-gluon interactions 
generate a hard $\sim 1/k_\perp^2$ tail for \mbox{TMDAs.} 
The basic elements of generating hard tails from soft primordial 
TMDAs are illustrated on scalar examples in 
Sect. \ref{Hard}.  Our conclusions  and directions of further applications 
of the VDA approach are discussed in Sect. \ref{Summary}.

\section{Transition form factor  in scalar model}
\label{Scalar}

\subsection{Choosing  representation}

\begin{figure}[h]
  \centerline{\includegraphics[width=2.2in]{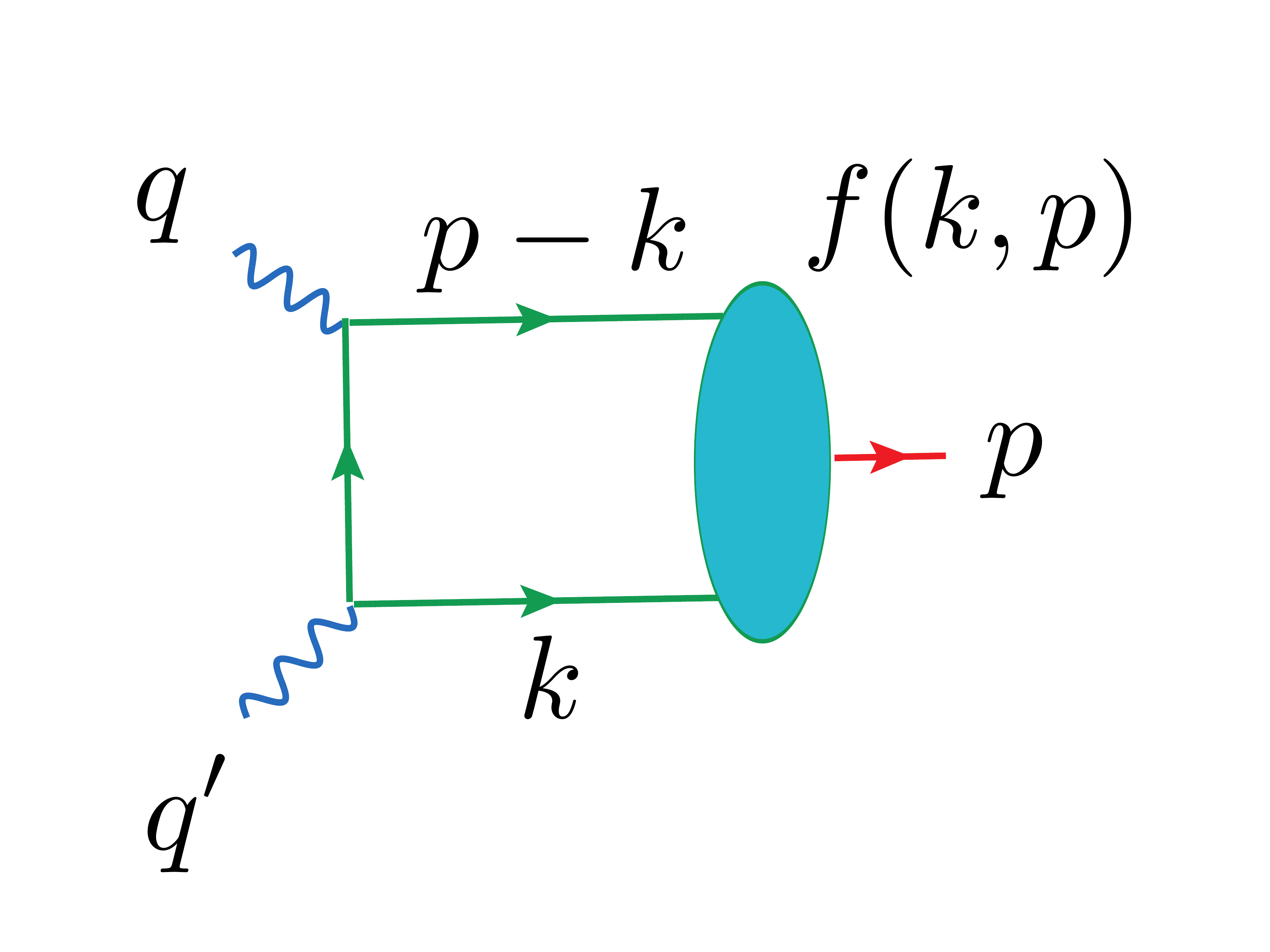}}
  \vspace{-0.5cm}
  \caption{General handbag diagram for photon-pion transition form factor in momentum representation.
  \label{sudak}}
  \end{figure}

We start with analysis of   general features of the handbag contribution to the
$\gamma^* \gamma \to \pi^0$ form factor (see Fig. \ref{sudak}).

In the momentum representation, the hadron structure is described by 
the hadron-parton   blob $f(k,p)$, which by  Lorentz invariance 
depends on 3  variables:   two parton virtuaities $k^2$, $(p-k)^2$, 
and the invariant mass  $p^2$, which   is $k$-independent.
None of  these invariants is convenient for extraction 
of the basic   parton  
variable  $x$ which is usually defined as the ratio $k^+ /p^+$ of 
the  ``plus''  light cone   components.

A standard way to  extract $k^+$   is to incorporate  a light-like vector
$n\equiv n^-$, which is additional to  variables  $k,p$
of the hadron-parton blob. Using 
 the Sudakov parametrization  $k= xp + \eta n + k_\perp$   \cite{Sudakov:1954sw}  for the $\gamma \gamma^* \pi^0$
amplitude,   it is convenient to  take $n=q'$, the momentum of the real photon.
Another convention was  used  in the light-front approach 
of Ref. \cite{Lepage:1980fj}, where $n$ is not directly related to the  
momenta involved in the process (in particular, $q' = p +    n \, q_\bot^2/(2 pn)- q_\bot$ 
in that definition).

\begin{figure}[t]
\centerline{\includegraphics[width=2in]{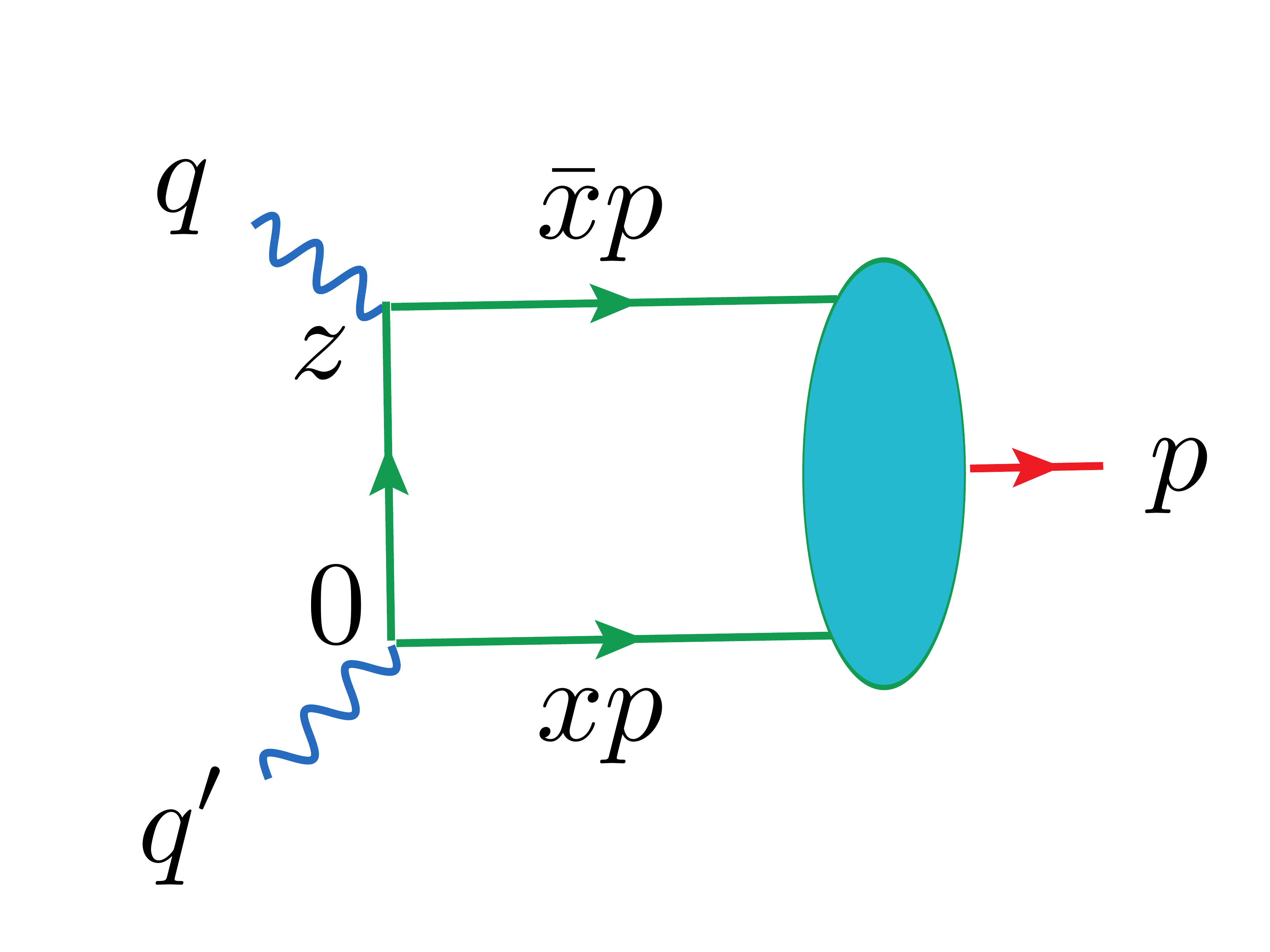}}
\caption{Handbag diagram in the coordinate  representation and parton momentum assignment.
\label{coord}}
\end{figure}

Switching to  the  coordinate representation, one  deals with 
the variable $z$ that is Fourier conjugate
to $k$ (see Fig.\ref{coord}).  The  blob  now  depends on $p$
and $z$, and   by Lorentz invariance is a function of  $(pz)$ and $z^2$.
Then  the parton fraction  $x$ may be defined   just as 
a variable that is Fourier conjugate
to $(pz)$. There is no need to have an external    vector 
like $n$ in such a definition, which is truly  process-independent
and involves only a minimal set of vectors $p,z$ 
describing the hadron state under  study.

In  
  the coordinate representation  (see Fig. \ref{coord}) we have 
  \begin{align}
T(p,q) = 
\int {d^4z}\, e^{-i(qz)} \, D^c(z) \, 
\langle  p | \phi (0)  \phi(z)  | 0 \rangle  
\label{eq:Fscalar}
\end{align}
   for a scalar handbag diagram,  where  $D^c(z) = -i/4\pi^2 z^2$ is  the scalar massless propagator,
$q$ is the momentum of the initial virtual  ``photon''
(\mbox{$q^2  \equiv - Q^2$}) 
given by $q=p-q'$, with 
$p$ being  the momentum of the final ``pion''.


  \subsection{Twist decomposition  of the bilocal operator}

The pion structure is described by the 
matrix element 
$\langle  p | \phi (0)  \phi(z)  | 0 \rangle $.
   To parametrize  it,  one may to wish to    
start with 
 the Taylor expansion 
 \begin{align}
  \phi(z) = \sum_{n=0}^{\infty} 
 \frac{1}{n!} \, z_{\mu_1} \ldots z_{\mu_n} \ 
 {\partial}^{\mu_1} \ldots
{\partial}^{\mu_n}   \phi(0) 
\ . 
\label{Taylor}
\end{align}
The next step is to  to write  
the tensor $z_{\mu_1} \ldots z_{\mu_n}$
 as a sum of products of powers of $z^2$ and 
symmetric-traceless
combinations $\{\ldots z_{\mu_i } \ldots z_{\mu_j} \ldots \}$ 
satisfying \mbox{$g^{\mu_i  \mu_j} 
\{\ldots z_{\mu_i } \ldots z_{\mu_j} \ldots\}=0$}.  Using the notation
$
\{z\partial\}^{n} \equiv  \{z_{\mu_1} \ldots z_{\mu_n} \}  \, 
\partial^{\mu_1} \ldots \partial^{\mu_n} 
$ 
for products of traceless tensors, we obtain 
 \begin{align}
 \label{expandphi}
  \phi(z) = \sum_{l=0}^{\infty} 
 \left ( \frac{z^2 }{4}  \right )^l  \sum_{N=0}^{\infty} 
 \frac{N+1}{l!(N+l+1)!}  
\{z{\partial} \}^{N}
 ({\partial}^2)^l 
 \phi(0)  . 
\end{align}
The operators containing powers of  ${\partial}^2$ have higher twist,
and  their contribution to the light-cone expansion
 is accompanied by powers of $z^2$.
Considering  the lowest-twist 
term, one   can write   matrix elements
of the local operators 
\begin{align} 
& \langle  p |\phi (0) \{  {\partial}^{\mu_1} \ldots
{\partial}^{\mu_n}  \}  \phi(0)  
| 0 \rangle 
= i^n 
\sum_{k=0}^{n} A_{n}  \,  \{ 
p^{\mu_1} \ldots  \ldots p^{\mu_{n}}   \} 
\end{align} 
in terms of  the  coefficients $A_{n} $. 
To perform summation over  $n$, 
one  can 
introduce  the \mbox{twist-2}   {\it distribution amplitude}  (DA)  $\varphi (x)$ 
as a function whose $\bar x^{n} $ moments 
($\bar x \equiv 1-x$)
are related  to  $A_{n}$ by  
\begin{align} 
\int_{0}^1 \varphi (x)\,  \bar x^n  dx 
=
A_{n}
\ .
\end{align}
The calculation of  the sum
\begin{align} 
\sum_{n=0}^{\infty} i^n 
\frac{ \bar x^{n} }{n!}
 \, 
\{z_{\mu_1} \ldots z_{\mu_n} \}\,
p^{\mu_1} \ldots \ldots p^{\mu_{n}} 
\equiv 
\sum_{n=0}^{\infty} i^n 
\frac{ \bar x^{n} }{n!}
 \, 
\{zp \}^n \,
\end{align}
is complicated  by 
traceless  combinations $\{z_{\mu_1} \ldots z_{\mu_n} \}$.
To this end, one can use the inverse expansion 
$\{zp \}^{n}  = (zp)^n - \frac14 {(n-1)} \, z^2 p^2 (zp)^{n-2} + \ldots $
to obtain  the \mbox{ parameterization} 
\begin{align}
& \langle p |   \phi(0) \phi (z)|0 \rangle 
=  
\int_{0}^1  \varphi (x)\,  \,  e^{i \bar x (pz) } \, dx
 +  {\cal O} (z^2)  
\label{DDF}
\end{align}
 of the
matrix element. 
The plane wave factor $e^{i \bar x (pz) }$ 
has a natural  interpretation  that the parton created at point  $z$ carries the fraction $\bar xp$ 
of the pion momentum $p$.  
  
The $(z^2)^k$ terms brought in by the re-expansion of 
$\{zp \}^{n}$  are accompanied 
by $(p^2)^k$   factors. 
These terms  are purely kinematical, and 
 may be summed explicitly, leading to an analog of 
Nachtmann \cite{Nachtmann:1973mr,Georgi:1976ve} corrections. 
To concentrate 
on  dynamical effects, one may    take   
 $p^2=0$, in which case 
 $\{zp \}^{n}  = (zp)^n$, and summation over $n$ is   
 straightforward.

  \subsection{Virtuality distributions}

For  a light-like momentum $p$,  the $ {\cal O} (z^2) $ 
terms in Eq. (\ref{DDF})   only  come from the 
$(z^2)^l$ terms of the original expansion (\ref{expandphi}) for $\phi (z)$.
These terms are accompanied by matrix elements 
of higher-twist operators
\begin{align}
\langle  p |\phi (0) \{ z {\partial} \}^k  
\left ({\partial}^2  \right )^l \phi(0)  
| 0\rangle \equiv [i (zp)]^k \Lambda^{2l} A_{kl}  \ .
\end{align} 
  The derivation  above 
 assumes 
  that the matrix elements of operators 
 containing  
 high powers of ${\partial}^2$  are finite, with their size characterized  by 
  some scale $\Lambda$, which  
 has an obvious meaning  of typical  virtuality 
of parton fields inside the hadron. 

For $l=0$, the coefficients $A_{k0}\equiv A_k$ reduce to those defining the twist-2
DA   $\varphi (x)$.  In general,  for each particular $l$, we can
 define the coefficients $A_{kl}$
to be  proportional to  the $\bar  x^k$ 
moments of  appropriate functions
$\varphi_l (x)$, and  arrive at  parametrization 
\begin{align}
 \langle p |   \phi(0) \phi (z)|0  \rangle 
   = & \sum_{l=0}^\infty \left ( \frac{ \Lambda^2 z^2}{4} \right )^l \int_{0}^1 dx \, \varphi_l (x) 
  \, e^{i \bar x (pz) }  
   \nonumber  \\
  \equiv   &
 \int_{0}^1 dx \, B(x, z^2/4)
 \, e^{i \bar x (pz) }  \label{Phixb}
\end{align} 
 in terms of the {\it bilocal  function}  $B(x, z^2/4)$.

  In fact, 
using the $\alpha$-representation as outlined in
Refs. \cite{Radyushkin:1983wh,Radyushkin:1983ea,Radyushkin:1997ki}, 
it can be demonstrated  that the contribution of 
any Feynman diagram  to  $\langle  p | \phi (0)  \phi(z)  | 0 \rangle $  
can be represented as 
\begin{align}
\langle  p | \phi (0) & \phi(z)  | 0 \rangle =  i^{l} \, \frac{P({\rm c.c.})}{(4\pi i)^{Ld/2}}
\int_0^{\infty} \prod_{j=1}^l   d\alpha_{j} [A(\alpha)+B(\alpha)]^{-d/2}
\nonumber \\ & \times 
\exp \left \{ -i  \frac{z^2/4}{A (\alpha)+B(\alpha) } +i(pz) \frac{B(\alpha) }{A (\alpha)+B(\alpha)}
 \right \} 
\nonumber \\ & \times 
\exp \left \{ 
 i p^2 C(\alpha) 
- i  \sum_{j} \alpha_{j} (m_{j}^2- i\epsilon) \right \}   \ , 
\label{alphaT}
\end{align}
where 
$d$ is the space-time dimension,  ${P({\rm c.c.})}$ is the relevant
 product of the coupling constants, 
 $L$ is the number of loops of the diagram, $l$
is the number  of its
internal lines,  and   
$A(\alpha), B(\alpha), C(\alpha)$ 
are positive functions of the $\alpha_\sigma$-parameters 
of the diagram.   
 Thus, we can write the matrix 
 element in the form 
\begin{align}
 \langle p |   \phi(0) \phi (z)|0 \rangle 
= & 
\int_{0}^{\infty} d \sigma \int_{0}^1 dx\, 
 \Phi (x,\sigma) \,  \,  e^{i \bar x (pz) -i \sigma {(z^2-i \epsilon )}/{4}} \,  ,  \label{Phixs0}
\end{align} 
{\it without any assumptions}  about regularity of the $z^2 \to 0$ limit.
There is also no need to assume that $p^2=0$.

 In a formal Taylor expansion of  $\phi(0) \phi (z)$ above, the $z^2$ factors 
 are accompanied by $\partial^2$, thus 
the  variable $\sigma$  is  related  to $\partial^2$, i.e.,   parton 
virtuality.  For this reason, we will refer to   the 
 spectral function $\Phi (x,\sigma)$ as 
the   {\it  virtuality distribution amplitude} (VDA). 
One should be aware, though, that while 
the virtuality $k^2$ may be both positive and negative, 
$\sigma$ is a positive parameter.

The VDA $\Phi (x,\sigma)$  is related to  the  {bilocal  function}  $B(x, z^2/4)$ 
 by 
 a Laplace-type representation
  \begin{align}
  B(x, \beta) 
  =  \int_{0}^{\infty}  \, e^{-i \beta \sigma} 
   \Phi (x,\sigma) \, d \sigma \ . 
\end{align} 
According to 
Eq. (\ref{alphaT}),   $B(x, z^2/4)$ 
 is a function of $z^2-i \epsilon$.

  \subsection{Transverse  momentum distributions}

The parton picture implies  a frame in which   $p$ has 
no transverse component,   a large 
component in  the  ``plus'' direction 
and a small component in the ``minus'' direction,
 so as  $p \to p^+$ when  $p^2 \to 0$. 
In the latter  limit,  only the ``minus'' component $z^-$ is essential in the product
$(pz)$.
In general,  without assuming $p^2=0$, we can take 
 a space-like separation $z$ having 
 $z^-$ and $z_\perp$ components only
  (i.e., $z^+=0$),  and 
 introduce 
 the {\it impact parameter distribution amplitude} (IDA) 
 $  {\varphi } (x, z_\perp)  \equiv B(x, -z_\perp ^2 /4)$,
  \begin{align}
 \langle p |   \phi(0) \phi (z)|0 \rangle  |_{z^+=0, p_\perp =0} 
   = 
 \int_{0}^1 dx \,  {\varphi } (x, z_\perp)
 \, e^{i \bar x (pz^-) }  \  .  \label{Phixbb}
\end{align} 
Note  that $B(x, \beta)$ is defined both for 
positive and negative $\beta$, while ${\varphi } (x, z_\perp)$
corresponds to negative $\beta$ only. 

The IDA function  
 may be also treated as a Fourier transform 
\begin{align}
{\varphi } (x, z_\perp) = \int  {\Psi}(x, k_\perp ) \, e^{i (k_\perp z_\perp)}
\,  {d^2 k_\perp }    \label{impaf} 
\end{align}
of the {\it transverse momentum dependent
distribution  amplitude}   (TMDA)  ${\Psi} (x, k_\perp )$.
 TMDA   can  be  written     in terms of VDA  as
\begin{align}
{\Psi} (x, k_\perp ) =&   \frac{i }{\pi }
\int_{0}^{\infty} \frac{d \sigma }{\sigma} \, 
 \Phi (x,\sigma) \,  \,  
 e^{- i (k_\perp ^2-i \epsilon )/ \sigma} 
 \  .   \label{Phixs1tmd} 
\end{align} 
  Actually  TMDA 
depends on  $k_\perp^2$ only: ${\Psi} (x, k_\perp )=  \psi (x,k_\perp^2)/\pi$.

This relation  is    quite general  in the sense that it  holds even 
if   the  $z^2 \to 0 $  limit  for the matrix element 
$\langle  p | \phi (0)  \phi(z)  | 0 \rangle$ of the bilocal operator 
is singular. However, if this limit is regular,
then the coefficients of the Taylor series of 
$B(x, \beta)$  in $\beta$  are given by $\sigma$ moments 
of the VDA   $\Phi (x,\sigma) $, which,  in turn, are related to 
$k_\perp^2$ moments of  TMDA ${\Psi} (x, k_\perp )$, namely 
\begin{align}
\int  {\Psi } (x, k_\perp ) \, k_\perp^{2n} \, d^2 k_\perp =  
\frac{n!}{i^n} 
\int_{0}^{\infty} {\sigma}^n  \, 
 \Phi (x,\sigma) \, {d \sigma }  \  .
 \label{sigma_k}
\end{align}
For these moments  to be finite,  
$\Phi (x,\sigma)$ should vanish for large $\sigma$ faster than any power of 
$1/\sigma$.  The functions having this property will be sometimes   referred to as ``soft''.
The lowest moment
\begin{align} 
\int_{0}^{\infty}  \Phi (x,\sigma) \, d \sigma  = \varphi  (x)   
\label{Phix0}
\end{align} 
 of VDA    $\Phi (x,\sigma)$
gives the usual twist-2 distribution amplitude  $\varphi(x)$.  
The reduction  relation for TMDA  ${\Psi} (x, k_\perp )$ 
\begin{align}
\label{redrel}
\int  {\Psi} (x, k_\perp ) \, d^2 k_\perp =  \varphi (x) 
\end{align}
is equivalent  to the reduction relation 
$ 
{\varphi } (x, z_\perp=0) = \varphi (x) 
$ 
 for the  IDA $ {\varphi } (x, z_\perp)$. 
 Using Eq. (\ref{sigma_k}), we derive 
\begin{align}
B(x, \beta) = \int  \Psi (x, k_\perp ) \, J_0 (2k_\perp  \sqrt{-\beta})\, d^2 k_\perp \ .
\label{zipsi}
\end{align}
For 
 negative   $\beta=-z_\perp^2/4$, this formula may be 
 also  obtained by performing the angular 
 integration in 
 \mbox{Eq. (\ref{impaf}).}  
If  \mbox{$\beta $}    is positive  (then we can write
$\beta= |z|^2/4$), one may understand  Eq.(\ref{zipsi}) as  
\begin{align}
B(x, |z|^2/4)  = \int  \Psi (x, k_\perp ) \, I_0 (k_\perp |z|)\, d^2 k_\perp \ ,
\label{tipsi}
\end{align}
 where $I_0$ 
is  the modified Bessel function.  Thus, in some cases, 
the   bilocal  function $B(x, \beta)$ may be 
 expressed in terms of TMDA
$ \Psi (x, k_\perp)$ both for spacelike and timelike values of $\beta$.
This fact  suggests that it might be  sufficient to know  just 
 the ``spacelike'' function  $ \Psi (x, k_\perp )$
to describe all  virtuality effects.

In our actual calculations,   
we have   no need to 
separate integrations over spacelike and timelike $z$
and use Eqs. (\ref{zipsi}), (\ref{tipsi}) and/or the presumptions 
(like Taylor expansion in $z^2$) 
on which they are based. 
All the coordinate  
integrations are   done  covariantly,  and furthermore, without any  
assumptions   whether   the $z^2\to 0 $ limit 
for $B(x, z^2/4)$ is regular or not.  The results are   expressed 
in terms  of VDA  $\Phi (x,\sigma)$, and then  
Eq.  (\ref{Phixs1tmd}) is     incorporated  that relates 
 $\Phi (x,\sigma)$   to the  TMDA  $ \Psi (x, k_\perp )$.
Our final   expressions are given in terms of TMDA.

\subsection{Handbag diagram in VDA representation}

The strategy outlined above may be 
illustrated by calculation of the Compton amplitude.
The starting expression   is given by  
    \begin{align}
T(p,q) =  \int_{0}^1 dx
\int {d^4z}\, e^{i(q'z)-i x (pz) } \, D^c(z) \, 
 \, B(x, z^2/4) \  . 
\label{eq:Fscalar1}
\end{align}
When the partons have zero virtuality,
i.e.  if all $\sigma^l$   moments with $l\geq 1$ vanish, 
we have $B (x, z^2/4) \Rightarrow  \varphi (x)$.
In this case, only the twist-2 operators contribute, and we have 
 \begin{align}
T(p,q)|_{\rm twist \ 2}  =& - \int_{0}^1 \frac{\varphi(x)\, dx}{(q'-xp)^2 +i \epsilon} 
 \nonumber \\ & =
 \int_{0}^1 \frac{\varphi(x) \, dx}{xQ^2 +x \bar x p^2 - i \epsilon}   \  . 
\label{eq:Fscalar}
\end{align}
The $x \bar x p^2$ term here produces   
target-mass 
 corrections analogous to those analyzed in Refs. \cite{Nachtmann:1973mr,Georgi:1976ve}. 
 For positive $Q^2$ and $p^2$, the denominator has no singularities, and
 $i \epsilon$ may be omitted. Also, given the smallness of the pion mass,
 we will neglect $p^2$ in what follows.  If necessary,  the target 
 mass corrections may be reconstructed  by  the appropriately made  changes  
 $xQ^2 \to xQ^2 + x \bar x m_\pi^2$ in the formulas given below.

To include the virtuality corrections,
we use  the  VDA parametrization (\ref{Phixs0})  that gives  
 \begin{align}
T(Q^2) = &
 \int_{0}^1 \frac{dx}{xQ^2} \, \int_{0}^{\infty}   d \sigma \, 
  { \Phi (x,\sigma)    } 
\left \{ 1- e^{-[ixQ^2 + \epsilon]/  \sigma }  \right \} \  . 
\label{eq:Fscalar3}
\end{align} 
The first term in the brackets does not depend on $\sigma$ and 
produces   the 
twist-2 expression (\ref{eq:Fscalar}). 
Using the TMDA/VDA relation (\ref{Phixs1tmd}), one may rewrite
Eq. (\ref{eq:Fscalar3})   in terms of TMDA as 
 \begin{align}
 T(Q^2) 
= &  
   \int_{0}^1   \frac{dx}{xQ^2}   \int_{k_\perp^2 \leq {x} Q^2}    \Psi (x, { k}_\perp   ) 
 \, d^2 { k}_\perp 
      \  . 
 \label{Phixs06}
 \end{align} 
 
\subsection{ TMDA and light-front wave function} 

One could notice that  the   definition of  TMDA  
  \begin{align}
 \langle p |   \phi(0) \phi (z)|0 \rangle &  |_{z_+=0, p_\perp =0} 
   = 
 \int_{0}^1 dx \, 
 \, e^{i \bar x (pz_-) }  \nonumber \\ & \times 
  \int  {\Psi}(x, k_\perp ) \, e^{i (k_\perp z_\perp)}
\,  {d^2 k_\perp } 
\label{tmda}
 \end{align} 
is similar to  that for  the  light-front wave function
used in Ref.  \cite{Lepage:1980fj}.  
The  difference is that  our approach 
is based on covariant calculations, with 
${\Psi}(x, k_\perp )$ appearing  at final stages
through its relation (\ref{Phixs1tmd}) to  a  covariant function
$\Phi (x, \sigma)$ which  is a generalization
of the DA  $\varphi (x)$. 
For this reason, 
 we shall continue to use the ``distribution amplitude''
terminology for the VDA-related functions.
In particular, ${\Psi}(x, k_\perp ) $ will be referred to as TMDA.

\section{Spin-1/2 quarks}
\label{Spin}

\subsection{Non-gauge case} 

When quarks have spin 1/2, the 
handbag  diagram for the pion transition form factor 
 is   given by 
 \begin{align}
\int {d^4z}\, e^{-i(qz)} 
  \langle p |   \bar \psi(0)  & \gamma^\nu \, S^c (-z) \, \gamma^\mu \,  \psi (z)|0 \rangle 
=i \epsilon^{\mu \nu \alpha \beta} p_\alpha q_\beta F(Q^2) \ , 
\label{fermiC}
\end{align}
where $S^c (z) = \slashed z/ 2\pi^2 (z^2)^2$ 
is the  propagator for a massless fermion.
Using that  the antisymmetric   part of $\gamma^\nu \, \slashed z \, \gamma^\mu $
is $i  z_\beta \epsilon^{\mu \nu \alpha \beta} \gamma_5 \gamma_\alpha $
and 
writing the twist-2 part of the matrix element 
 \begin{align}
&  \langle p |   \bar \psi(0) \gamma_5 \gamma_\alpha   \, \psi (z)|0 \rangle  |_{\rm twist \ 2}  
\ 
\label{pmurmu2}
 =i p_\alpha
\int_0^1 dx \,   e^{i \bar x (pz) } \,
\, \varphi (x) 
\end{align}
one obtains the same   formula
 \begin{align}
F(Q^2) |_{\rm twist \ 2}  =  \int_{0}^1 \frac{\varphi (x)}{xQ^2} \,  dx  
\label{eq:Fspinor}
\end{align}
as in the scalar model considered above. 
The higher-twist operators  may be included 
 through the  VDA parametrization 
\begin{align}
  \langle p |   \bar \psi(0)   \,\gamma_5 \gamma_\alpha   \,  \psi (z)|0 \rangle  
 & = i   p_\alpha
\int_{0}^{\infty} d \sigma \int_{0}^1 dx\,  \nonumber \\ & \times 
 \Phi (x,\sigma) \,  \,  e^{i \bar x (pz) -i \sigma {(z^2-i \epsilon )}/{4}} 
 + \ldots  \ , \label{Phixspin}
\end{align} 
in which we  omitted  the  terms proportional
to $z_\alpha$, since they disappear after 
convolution with  $z_\beta\epsilon^{\mu \nu \alpha \beta} $. All  the formulas relating    VDA with  IDA and TMDA  that
 were   derived in the  scalar case are valid   without changes.
However, the spinor propagator has a different $\sim \slashed z/ (z^2)^2$ functional form.
Using  it, we obtain 
 \begin{align}
F(Q^2) = &
\int_{0}^{\infty}   d \sigma  \int_{0}^1 { \Phi (x,\sigma)    } \frac{dx}{xQ^2} \, 
\left \{ 1+ \frac{i\sigma}{xQ^2} \left [ 1- e^{-[ixQ^2 + \epsilon]/  \sigma } \right ] \right \} \  . 
\label{eq:Fspinor3}
\end{align} 

For large $Q^2$,  Eq.  (\ref{eq:Fspinor3})  contains  a power-like  
$\sim1/Q^4$ correction    that 
corresponds to the  twist-4 
$\bar \psi \gamma_5 \gamma_\alpha 
D^2 \psi $ operators.  Though they are  accompanied by 
a $z^2$ factor,  the latter does not completely 
cancel the $1/z^4$ singularity
of the  spinor propagator,    and 
 this contribution has a ``visible'' $ \Lambda^2/Q^4$ 
 behavior\footnote{The scale $\Lambda^2$ is set by
$\langle \pi |\bar \psi \gamma_5 \gamma_\alpha 
D^2 \psi |0 \rangle =  \Lambda^2 \langle \pi |\bar \psi \gamma_5 \gamma_\alpha 
\psi |0 \rangle$, with $\Lambda^2 =0.2$\, GeV$^2$ \cite{Novikov:1983jt}
 being a widely accepted value.}.  
The remaining term contains 
contributions ``invisible'' in the OPE.
In terms of TMDA we have 
  \begin{align}
 F(Q^2) 
= &  
   \int_{0}^1   \frac{dx}{xQ^2}   \int_{0}^{xQ^2}  \frac{d k_\perp^2}{ xQ^2} 
    \int_{{k'_\perp}^2 \leq  k_\perp^2}   \Psi (x,  k'_\perp   ) 
 \, d^2  k'_\perp
      \  . 
 \label{Fspinor40}
 \end{align}
 
 With the help of Eq. (\ref{Fspinor40}), one can  
calculate  form factor in various models for   the TMDA 
$ \Psi (x, { \kappa}_\perp   ) $.
Before analyzing  form factor in some simple models for
 VDA, 
we show in the next section   that  switching to gauge theories
 does not bring any  further changes in general equations.


\subsection{Gauge theories }

\begin{figure}[t]
   \centerline{\includegraphics[width=2.5in]{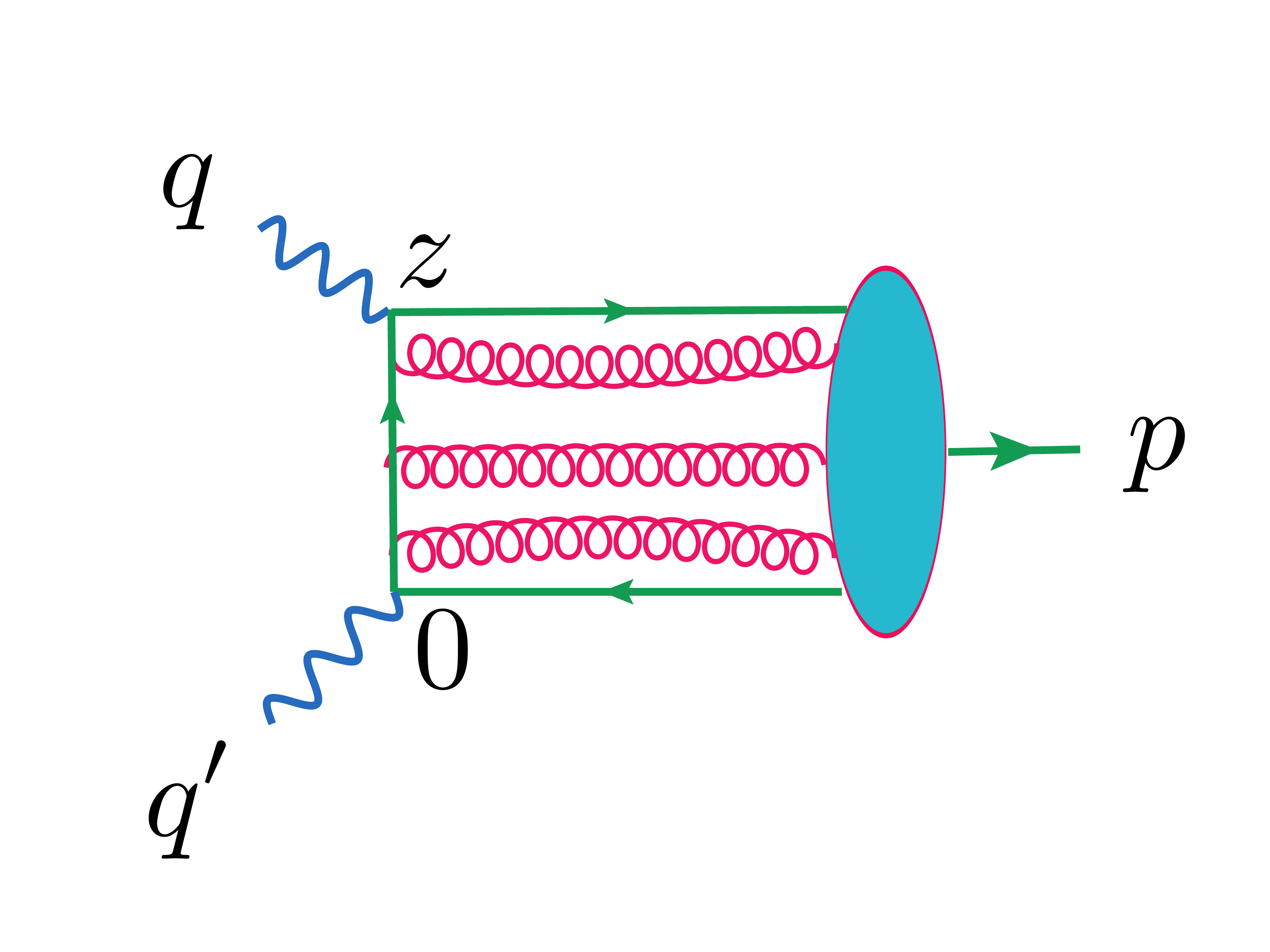}}
   \caption{Structure of the  handbag contribution in QCD.
   \label{QCD}}
   \end{figure}

In gauge theories, the handbag  contribution in 
a covariant gauge should be complemented 
by   diagrams corresponding to operators 
$\bar \psi (0)  \ldots A(z_1) \ldots  \psi (z)$
containing twist-0 gluonic field $A_\mu (z_1)$ inserted 
into the fermion line between the points 0 and $z$
(see Fig.\,\ref{QCD}).
It is well known \cite{Efremov:1978fi,Efremov:1978xm,Efremov:1980ub}  
that  these insertions may be organized to produce  
a  path-ordered exponential
\begin{align}
{ E}(0,z; A) \equiv P \exp{ \left [ ig \,  z_\mu\, \int_0^1dt \,   A^\mu (t z) 
 \right ] } 
 \label{straightE}
\end{align}
accumulating the zero-twist field $A^\mu$, and 
insertions of the non-zero twist gluon field
\begin{align}
 {\mathfrak A}^\mu  (z) = z_\nu 
 \int_0^1 \,   G^{\mu \nu} (s z) \, s \, ds \ ,
 \label{FSA}
\end{align}
which is the 
vector potential in the Fock-Schwinger 
gauge  \mbox{\cite{Fock:1937aa,PhysRev.82.664}.}
These insertions correspond to three-body $\bar qG q$, etc. 
higher Fock components.  At the   two-body  $\bar q q$ Fock component  level,  
 we  deal with the gauge-invariant bilocal operator
\begin{align}
{\cal O}^\alpha  (0,z; A) \equiv \bar \psi (0) \,
 \gamma_5 \gamma^\alpha \,  { E} (0,z; A) \psi (z)  \  .
\end{align}
An important property of this operator  
 is that the   Taylor expansion for  
 ${\cal O}^\alpha  (0,z; A) $   has precisely 
the same structure 
as that for the original $\bar \psi (0)  \gamma_5 \gamma^\alpha \psi (z)$ 
operator, with the only change that
one should use covariant derivatives 
\mbox{$D^\mu =\partial^\mu  - ig A^\mu$}  instead of the
ordinary  $\partial^\mu $ ones:
\begin{align}
  { E} (0,z; A)\,  \psi (z) = \sum_{n=0}^{\infty} 
 \frac{1}{n!}\, z_{\mu_1} \ldots z_{\mu_n} \ 
 {D}^{\mu_1} \ldots
{D}^{\mu_n}  \psi (0) \ . 
\label{Taylor}
\end{align}
This means that one can 
introduce the  parametrization 
\begin{align}
  \langle p | \, {\cal O}^\alpha  (0,z; A)  |0 \rangle  
  = & i p^\alpha 
\int_{0}^{\infty} d \sigma \int_{0}^1 dx\,  
 \Phi (x,\sigma) \,  \,  e^{i \bar x (pz) -i \sigma {(z^2-i \epsilon )}/{4}} 
 \nonumber \\ & + z^\alpha \ldots 
 \label{OPhixspin0}
\end{align} 
that accumulates information about   higher twist terms in VDA
$\Phi (x,\sigma)$
and  proceed exactly like in a non-gauge case.

\section{ Modeling soft  transverse momentum dependence }
\label{TMDA_Models}

To analyze  predictions of  the 
 VDA  approach  for
 $F(Q^2)$,   we will use  Eq.  (\ref{Fspinor40}) 
incorporating there    particular  models for  VDA $\Phi (x, \sigma)$
or  TMDA $\Psi (x, k_\perp)$.
For  illustration, we consider   first     the  simplest  case  of  factorized models
\mbox{$\Phi (x, \sigma) =\varphi (x) \,  \Phi (\sigma)$}  or 
  \begin{align}
\Psi (x, k_\perp) = \varphi (x) \, \psi (k_\perp^2)  \ ,
\end{align} 
  in which the  $x$-dependence and  $k_\perp$-dependence appear in  separate factors,
  and later  consider some non-factorized  models.

\subsection{Gaussian model} 

It is  popular to assume  a Gaussian dependence on
$k_\perp$, 
  \begin{align}
\Psi_G (x, k_\perp) = \frac{\varphi (x)}{\pi \Lambda^2}  e^{-k_\perp^2/\Lambda^2} \ . 
\label{Gaussian}
\end{align} 
In the impact parameter space, one gets IDA
  \begin{align}
\varphi_G (x, z_\perp) ={\varphi (x)} \, e^{-z_\perp^2 \Lambda^2/4} 
\label{GaussianI}
\end{align} 
that also  has a Gaussian dependence on $z_\perp$.
Writing 
\begin{align}
 \varphi_G (x, z_\perp)=&\frac{ {\varphi (x)} }{2\pi  }
\int_{-\infty}^{\infty} \frac{ i  \, d \sigma }{\sigma +i \Lambda^2} \, 
 \,  
 e^{- i z_\perp ^2 \sigma/4} 
 \  ,  \label{gausssigma} 
\end{align} 
we see that the integral involves both positive and
negative $\sigma$, i.e. formally  $\varphi_G (x, z_\perp)$ 
cannot be written in the VDA representation 
(\ref{Phixs1tmd}). However, the form factor formula in terms of TMDA 
(\ref{Fspinor40})  shows no peculiarities in case of the Gaussian ansatz,
so we will  use  this model  because of its calculational simplicity.

\subsection{Simple non-Gaussian models}

 One may  also argue that  the  Gaussian 
 ansatz (\ref{GaussianI}) has too  fast a 
 fall-off for large $z_\perp$.  
For comparison, 
 the 
 propagator  
 \begin{align} 
D^c(z,m)
  =   
 \frac1{16 \pi^2}  \int_0^{\infty} e^{-i \sigma z^2/4 - i  (m^2 - i\epsilon)/\sigma  }  
 {d \sigma}  
\label{alpharD}
\end{align}
 of a massive particle   
 falls off as  $\sim e^{-|z| m}$
for large  space-like 
distances.  
At small $z^2$, however,  the free particle propagator 
has $1/z^2$ singularity while we want 
a model for $\langle p| \phi (0) \phi (z) |0 \rangle$ that is  finite 
at $z=0$. The simplest way is to add a constant term 
$(-4/\Lambda^2)$ to $z^2$ 
in the VDA representation  (\ref{Phixs0}). So, we take
\begin{align} 
\Phi (x, \sigma) = \frac{\varphi (x)}{ p(\Lambda,m)} e^{i \sigma /\Lambda^2  - i  m^2 /\sigma -\epsilon \sigma }  
\label{alpharD}
\end{align}
as a model for VDA. 
The sign of the $\Lambda^2$ term is fixed 
from the requirement that $(4/\Lambda^2-z^2)^{-1}$ should not 
have singularities for space-like $z^2$.  The normalization factor  $p(\Lambda,m)$
is given by
\begin{align} 
{ p(\Lambda,m)} = \int_0^\infty e^{i \sigma /\Lambda^2  - i  m^2 /\sigma   -\epsilon \sigma }  d \sigma =
2i {m}{\Lambda} K_1 (2 m/ \Lambda)  \ .
\label{alpharD}
\end{align}

\subsection{$m=0$ models} 
\label{secm0}

To concentrate  on the effects of introducing $\Lambda$, 
 let us take  $m=0$, i.e. consider 
\begin{align} 
\Phi_s (x, \sigma) = {\varphi (x)} \,  \frac{e^{i \sigma /\Lambda^2  
  -\epsilon \sigma} }{ i \Lambda^2 } \ .
\label{alpharD0}
\end{align}
The bilocal matrix element in this case is given by
\begin{align}
& \langle p |   \phi(0) \phi (z)|0 \rangle 
=  
\, \frac{1}{1-z^2 \Lambda^2 /4}
\int_{0}^1 dx\, {\varphi (x)} \,  e^{i x (pz)}  \,  ,
\label{Phixs}
\end{align} 
which  corresponds to 
\begin{align}
{\varphi }_s (x, z_\perp) =
\, \frac{\varphi (x)}{1+ z_\perp^2   \Lambda^2/4 }  
\label{impaxs}
\end{align} 
for  IDA.  
Note  that the $z_\perp^2$ term of the 
$z_\perp$ expansion  of  ${\varphi } (x, z_\perp)$
in this model was adjusted to coincide with that of the exponential model,
so that    $\Lambda^2$ has the same 
meaning of  the  scale of $\phi \partial^2 \phi$  operator.
 The TMDA for this  ansatz is given by
\begin{align}
& {\Psi}_s (x, k_\perp ) = 2 \varphi (x)\,  \frac{K_0 ( 2 k_\perp / \Lambda) }{ \pi  \Lambda^2 } 
\  .
\end{align}
It has a logarithmic singularity for small $k_\perp$
that reflects a slow $\sim 1/z_\perp^2$ fall-off  of  ${\varphi}_s (x, z_\perp)$ 
for large  $z_\perp$.  
The integrated TMDA that enters the form factor formula
(\ref{Fspinor40})  is given by 
  \begin{align}
    \int_{{k'}_\perp^2 \leq  k_\perp^2}   \Psi (x, { k'}_\perp   ) 
 \, d^2  k'_\perp = 
 \varphi (x) 
  \left [ 1- \frac{2k_\perp }{ \Lambda}  K_1 ( 2 k_\perp / \Lambda) \right ]
      \  . 
 \label{Fspinor4}
 \end{align}
It is also possible to   calculate explicitly the next $k_\perp$ integral 
 involved there, see Eq. (\ref{eq:Fscalar3htmod12}) below.

\subsection{$m\neq 0$ model} 
\label{mne0}

The model with nonzero mass-like term 
\begin{align} 
\Phi_m (x, \sigma) = \varphi (x) \frac{e^{i \sigma /\Lambda^2  - i  m^2 /\sigma  } }{ 2i {m}{\Lambda} K_1 ( 2 m/ \Lambda)  }  
\label{alpharD2}
\end{align}
corresponds to the  function 
\begin{align}
& \Psi_m(x, k_\perp ) = \varphi (x)\,  \frac{K_0 \left (2 \sqrt{ k_\perp^2 +m^2 } / \Lambda \right ) }{ \pi  m \Lambda
 K_1 (2 m/ \Lambda) }  \  .
 \label{psim}
\end{align}
 that  is finite for $k_\perp =0$ in accordance with the  fact that
the IDA 
\begin{align}
{\varphi}_m (x, z_\perp)
=  {\varphi (x)}   \,  \frac{ K_1 \left ( m\sqrt{4/ \Lambda^2 + z_\perp^2} \, \right ) }{
 K_1 (2 m/ \Lambda) \, \sqrt{1 +\Lambda^2   z_\perp^2/4 }  }    
\end{align} 
in this case  has  $\sim e^{-m |z_\perp|}$
fall-off for  large $z_\perp$.  For small  $z_\perp$, we have 
${\varphi}_m (x, z_\perp) =  {\varphi (x)} [1- z_\perp^2  \tilde \Lambda^2 (\Lambda,m)
+\ldots ] $ behavior.

\section{Modeling transition form factor}
\label{FF_Model}

Let us now use these models to calculate the transition form factor
with the help of  Eq.(\ref{Fspinor40}). 

\subsection{Gaussian model}

In case of  the Gaussian model (\ref{Gaussian}), we have 
 \begin{align}
F_G (Q^2) = &
 \int_{0}^1 \frac{dx}{xQ^2}   \, 
{\varphi (x)} \, 
 \left [ 1-  \frac{\Lambda^2}{xQ^2}  \left (1 -   e^{-xQ^2/\Lambda^2} \right ) \right ]  \  .
\label{FGauss2}
\end{align} 
For large $Q^2$,  Eq.  (\ref{FGauss2}) 
 displays  the power-like twist-4 contribution and the 
 term that   falls faster than any power of $1/Q^2$. 
Note that  
the $x$-integral for the purely twist-2 contribution
converges if the pion DA $\varphi (x)$ vanishes 
as any positive power $x^\alpha$ for \mbox{ $x \to 0$,}
while the total integral 
in Eq. (\ref{FGauss2})  converges 
even for singular DAs  $\varphi (x) \sim x^{-1+\alpha}$ 
with arbitrarily  small $\alpha$. 
Furthermore,    the formal $Q^2 \to 0$ limit is finite: 
 \begin{align}
F_G (Q^2=  0) 
=  \frac{ f_\pi } { 2 \Lambda^2}    \  ,
\label{FGauss4}
\end{align} 
where we have used  the normalization condition 
 \begin{align}
 \int_{0}^1  \varphi (x) \, dx  =f_\pi  \  . 
\label{eq:phinorm}
\end{align} 
Note that $F(Q^2)$ is finite for $Q^2=0$ in any model with finite
$\Psi  (x, 
k_\perp=0)$. According to Eq. 
(\ref{Fspinor40}), one has then 
 \begin{align}
F (Q^2=  0) 
  = \frac{\pi}{2} \int_0^1 \Psi  (x, 
k_\perp=0) \, dx  \  .
\label{FQ0}
\end{align} 

\subsection{$m=0$ model} 

Using the  non-Gaussian model  (\ref{alpharD0})  with $m=0$  gives 
 \begin{align}
F (Q^2) = &
 \int_{0}^1  \frac{dx}{xQ^2} \, 
{\varphi (x)} \, 
\left [ 1-    \frac{\Lambda^2}{xQ^2}  + 2  
K_2 (2 \sqrt{x }Q/\Lambda) 
 \right ] 
\ . 
\label{eq:Fscalar3htmod12}
\end{align} 
 The size of the twist-4 term here 
is  given by the confinement scale
$\Lambda^2$,
just as in the Gaussian model.

\subsection{$m\neq 0$ model} 

Turning  to the $m \neq 0$ model, we have 
 \begin{align}
 F_m & (Q^2) = 
 \int_{0}^1  \frac{dx}{xQ^2} \, 
{\varphi (x)} \, 
\left \{1-  \frac1{ xQ ^2 } \frac{ \Lambda m }{ K_1 (2m/\Lambda) }  \right. 
\nonumber \\ & \left. \times \left [K_2 (2m/\Lambda)-  
  \left (1+\frac{x Q^2}{m^2}  \right ) 
  K_2 (2{\sqrt{x Q^2 +m^2}/\Lambda}) \right ]
 \right \} \ . 
\label{eq:Fspinor3htm}
\end{align}
Now, the size of   the twist-4 power 
correction depends   on the interplay of the 
confinement scale  $\Lambda$ and mass-type scale $m$.
 
 \subsection{Comparison with data}

 In QCD, the  twist-2  approximation 
 for    $F (Q^2)$  in the leading (zeroth) order in $\alpha_s$
 is 
  \begin{align}
F^{\rm LOpQCD} (Q^2)= & 
 \int_{0}^1  \frac{dx}{xQ^2} \, 
{\varphi (x)}  \ . 
\label{pQCD}
\end{align} 
 Taking the value of $I (Q^2) \equiv Q^2 F (Q^2)/f_\pi$ from the data 
 gives  information about the shape of the pion DA.  
 In particular,  for DAs of  $\varphi_r (x) \sim (x \bar x)^r$ type, one has  $I_r^{\rm LOpQCD}(Q^2) =1+2/r$,
i.e.   \mbox{$I^{\rm as}(Q^2)  = 3 $} 
 for the ``asymptotic'' wave function $\varphi^{\rm as} (x)  =  6 f_\pi x\bar x$.

\begin{figure}[h]
\centerline{\includegraphics[width=2.8in]{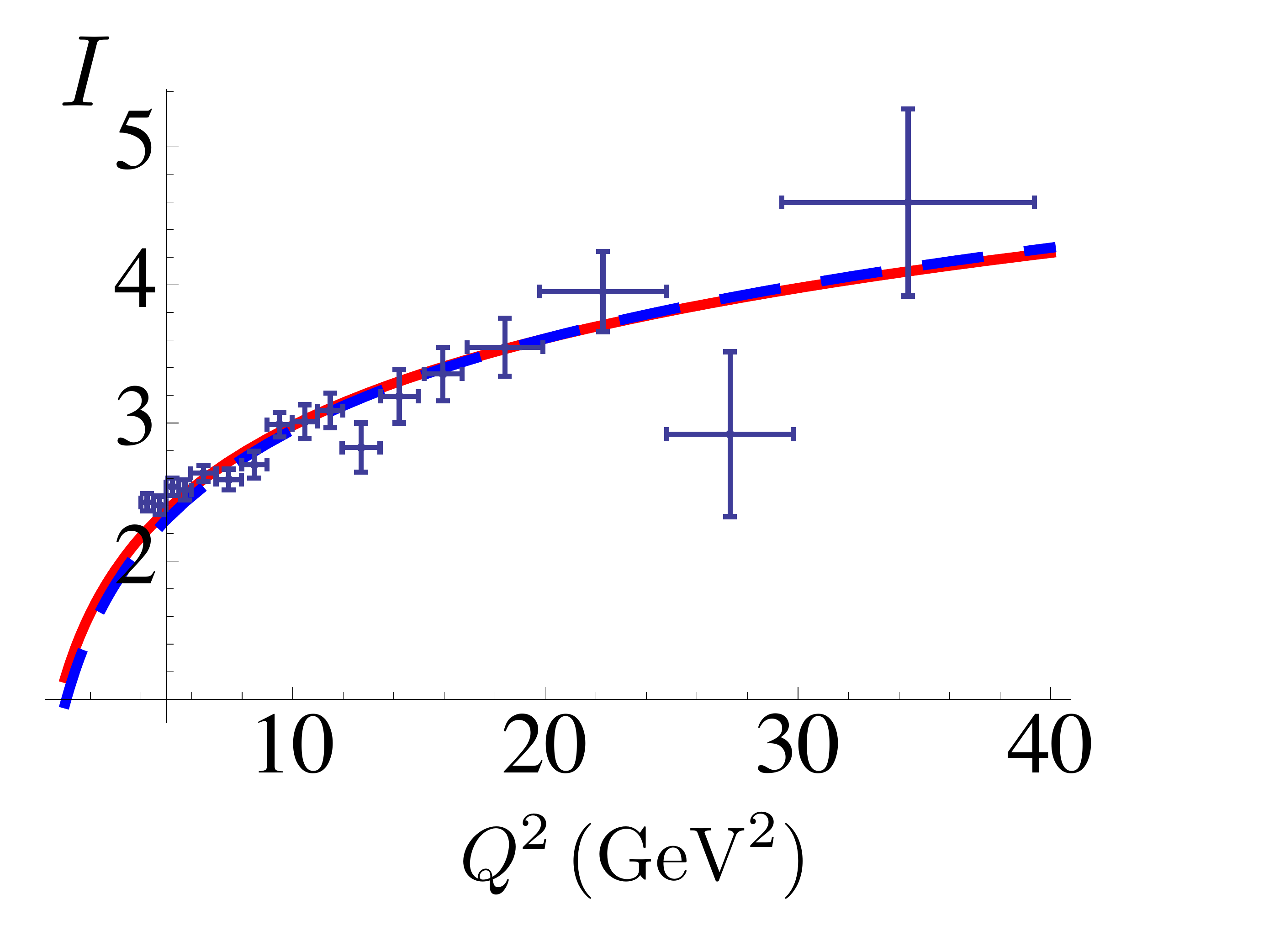}}
\vspace{-0.4cm}
\caption{BABAR data compared to model curves  described in the text.
\label{babar}}
\end{figure}

\begin{figure}[t]
\centerline{\includegraphics[width=3.1in]{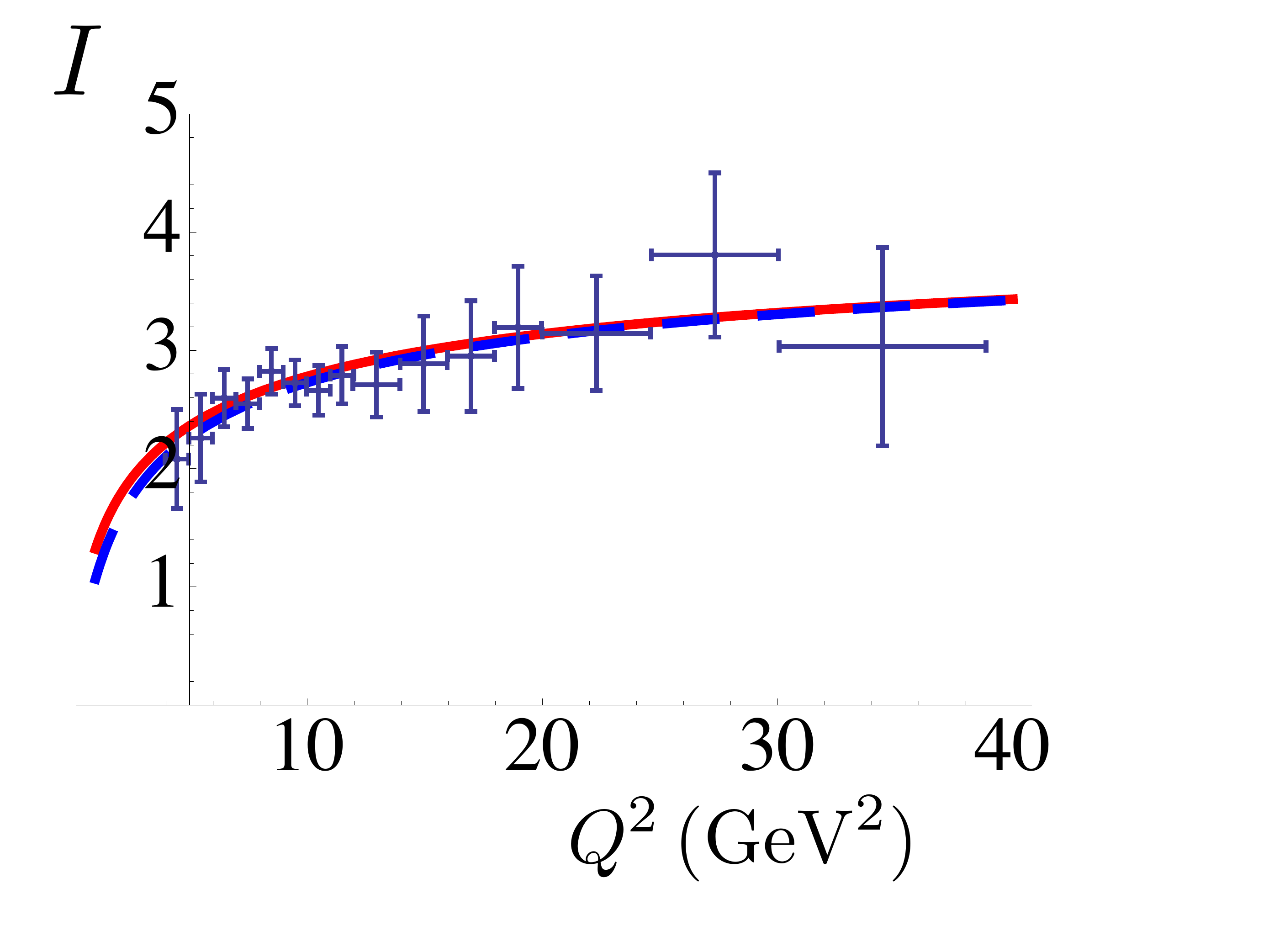}}
\vspace{-0.6cm}
\caption{BELLE data compared to model curves 
described in the  text.
\label{belle}}
\end{figure}

 The most  recent data \cite{Aubert:2009mc,Uehara:2012ag}
 still show a $Q^2$ variation of   $I(Q^2)$ (see Figs. \ref{babar}, \ref{belle}),
 especially in case of BaBar data \cite{Aubert:2009mc} which  contain several 
 points with $I(Q^2)$ values well above 3.  It was argued \cite{Radyushkin:2009zg,Polyakov:2009je} 
 that BaBar data indicate that the pion DA is close to a flat function $\varphi^{\rm flat}  (x) = f_\pi $.
 The latter corresponds to $r=0$, and pQCD gives $I^{\rm flat}=\infty$. 
 As shown in Ref. \cite{Radyushkin:2009zg}, inclusion of
  transverse momentum dependence
 of the pion wave function in the  light-front formula  
 of Ref.  \cite{Lepage:1980fj} (see also \cite{Musatov:1997pu})
   eliminates the divergence at $x=0$, and
 one 
 can produce a curve that fits the BaBar data. 
 Similar curves may be obtained within the VDA approach described in the present  paper.

In Fig.  \ref{babar}, we compare BaBar data with model curves 
corresponding to  flat DA  $\varphi (x) =f_\pi $ and two types of 
transverse momentum distributions. First, we take the 
Gaussian model 
of Eq.   (\ref{FGauss2}). 
A curve closely following the data is obtained for 
a value of $\Lambda^2=0.35\,$GeV$^2$  which is 
larger than the standard estimate $\Lambda^2=0.2\,$GeV$^2$ 
\cite{Novikov:1983jt} for the 
matrix element of the $\bar \psi \gamma_5 \gamma_\alpha D^2 \psi$ operator.
However, the higher-order pQCD corrections are known 
\cite{Li:1992nu}   to 
shrink the $z_\perp$ width  of the IDA $\varphi (x, z_\perp)$,
effectively increasing  the observed $\Lambda^2$ compared to the 
primordial value of  $\Lambda^2$.
For illustration, we also take the non-Gaussian $m=0$  model of Eq. (\ref{eq:Fscalar3htmod12}),
 to  check  what  happens 
 in  case of unrealistically slow  $\sim 1/z_\perp^2$ decrease 
  for large $z_\perp$.   Still, 
if we take  a   larger value of 
$\Lambda^2=0.6\,$GeV$^2$, this model  
 produces practically the same curve 
as the $\Lambda^2=0.35\,$GeV$^2$ Gaussian model. 

 Data from BELLE \cite{Uehara:2012ag} give lower values for $I$,
suggesting a non-flat DA. In Fig.  \ref{belle},
 we show the curves corresponding to  
\mbox{$\varphi (x) \sim f_\pi (x\bar x)^{0.4}$} DA.
If  we take the  
Gaussian model   (\ref{FGauss2}),  a good eye-ball fit to data is  produced
if we take 
$\Lambda^2=0.3\,$GeV$^2$.
Practically the same curve is obtained in  the   non-Gaussian
 \mbox{$m=0$}   model  
of Eq. (\ref{eq:Fscalar3htmod12})  for 
 $\Lambda^2=0.4\,$GeV$^2$. Again, a VDA-based analysis  of the higher-order 
 Sudakov effects \cite{Li:1992nu}   is needed to extract 
the value of $\Lambda$ in the primordial TMDA.

\section{Modeling hard tail}
\label{Hard} 

Higher-order pQCD corrections also modify the large-$k_\perp$ behavior 
of TMDA, producing a hard $\sim 1/k_\perp^2$ tail. 
Matching the soft and hard parts of transverse momentum distributions
is a very important problem in their studies.
 Below, on simple  scalar model examples  we illustrate
the basics  of  using the VDA approach for  
generation of  hard tail terms from original purely  soft distributions.

\subsection{Simple  model for   hard TMDA}

   \begin{figure}[b]
   \centerline{\includegraphics[width=1.6in]{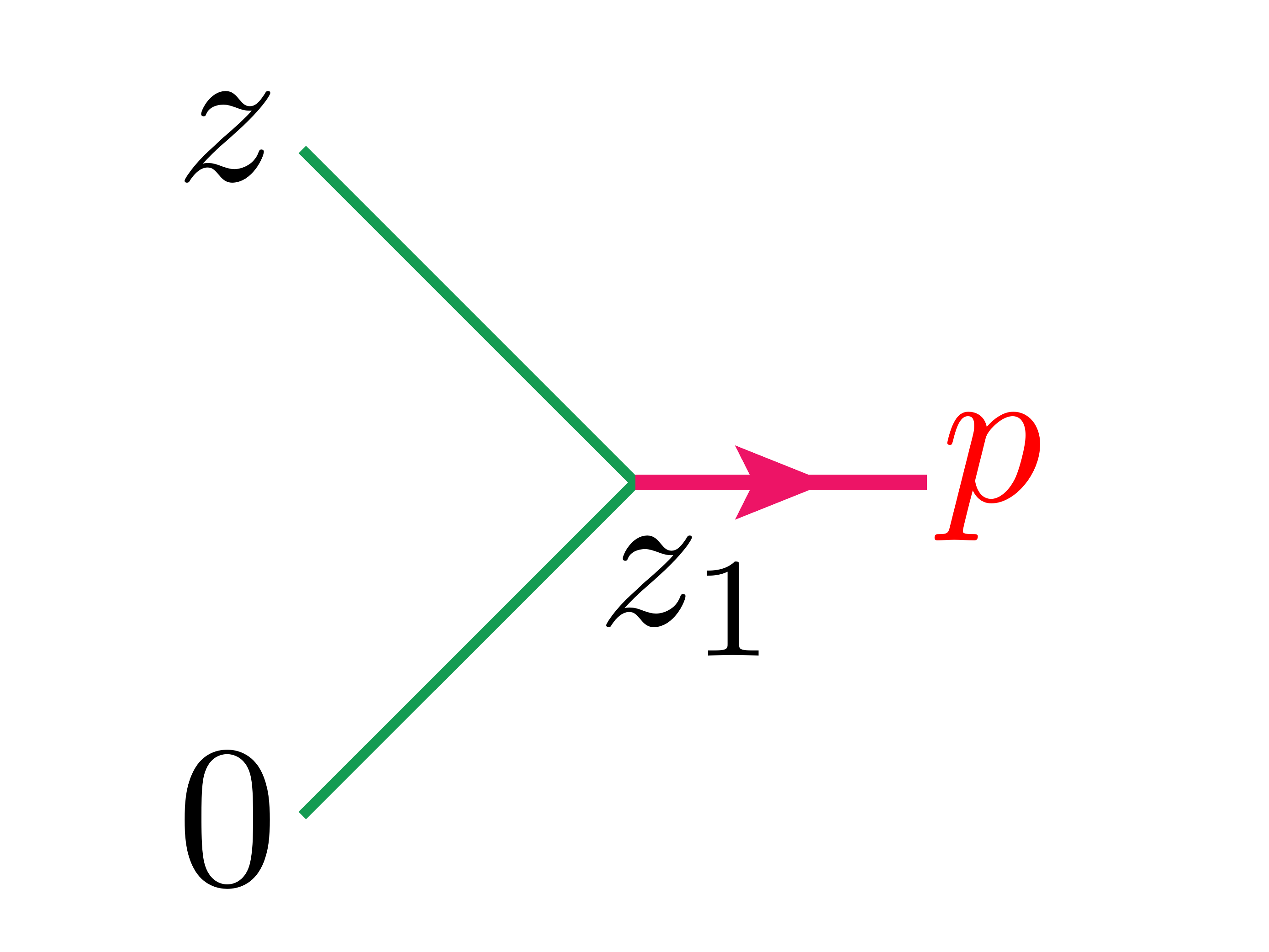}}
   \caption{Modeling VDA by a local current source.
   \label{current}}
   \end{figure}

Modeling the matrix element 
$ \langle p |   \phi(0) \phi (z)|0  \rangle$ 
 by two   propagators $D^c(z_1,m)$ and \mbox{$D^c (z-z_1,m)$}
 (see Fig. \ref{current}),
 with 
momentum $p$ going out of the point $z_1$  gives (after integration over $z_1$) 
  \begin{align}
\Phi^{\rm point} (x, \sigma) =   \frac1{\sigma} \, e^{i (x \bar x p^2 -m^2)/\sigma } 
\end{align} 
for the analog of VDA. For TMDA, this  yields  
  \begin{align}
\Psi^{\rm point} (x, k_\perp) =   \frac1{\pi} \, \frac1{k_\perp^2+m^2-x \bar x p^2  }  \ .
\label{psi2}
\end{align} 
It has a hard powerlike $1/k_\perp^2$ tail for large $k_\perp$. 
  Making a  formal $d^2 k_\perp$ integration  to 
 produce DA, one faces  in this case a logarithmic divergence.  In the impact parameter space, we  have 
 $\varphi_2 (x, z_\perp) = 2 K_0 (z_\perp \sqrt{m^2 - x \bar x p^2})$, a function with  a logarithmic
 singularity for $z_\perp =0$,  which is another manifestation 
 of the divergence of the $k_\perp$ integral for  
 $\Psi_2 (x, k_\perp) $.
In fact, the function $\Psi^{\rm point} (x, k_\perp)$   has a ``bound state'' pole 
in $p^2$ at 
the location given by 
 a well-known  light-front combination
$(k_\perp^2+m^2)/x\bar x$.  
However, we see no reasons to expect that 
in general TMDAs $\Psi (x, k_\perp)$ 
depend  on $k_\perp$ through $k_\perp^2/x\bar x$.

\subsection{Hard  exchange model}

\begin{figure}[h]
   \centerline{\includegraphics[width=1.6in]{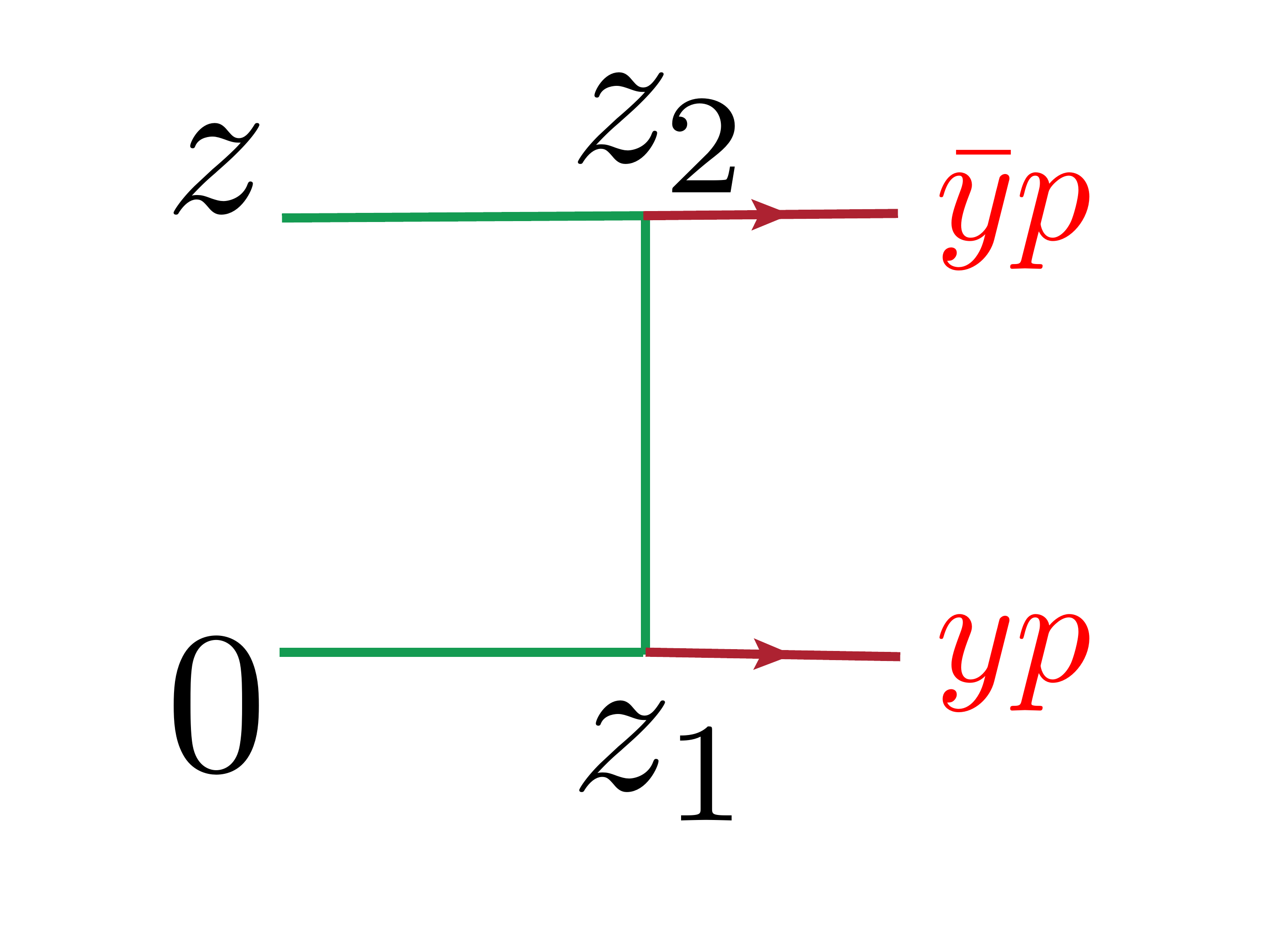}}
   \caption{Modeling VDA by a two-current state.
   \label{twocurrent}}
   \end{figure}

 A more complicated toy model involves two  currents 
 carrying momenta 
 $yp$ and $ (1-y)p   \equiv  \bar y p$
 at locations $z_1$ and $z_2$, respectively  (see Fig. \ref{twocurrent}).
 With  an exchange interaction
 described by a scalar propagator
 $D^c (z_1-z_2,m)$, 
 we have
  \begin{align}
\Phi^{\rm 2-curr} (x, \sigma;y) =&   i g^2 \,\frac{e^{i (x \bar x p^2 -m^2)/\sigma }  }{16\pi^2 \, \sigma^2 } \, 
 \int_0^{{\rm min} \left \{\frac{x}{y}, \, \frac{\bar x}{\bar y} \right \}}   e^{ -i 
 y \bar y\beta\,  {p^2} /{\sigma }  }
\, d \beta
\end{align} 
as  an analog of VDA,   where $g$ is the coupling constant.
For $p^2=0$, the $\beta$-integral gives   a well-known combination 
  \begin{align}
V (x, y) = \frac{x}{y} \, \theta(x<y) +  \frac{ \bar x}{\bar y } \, \theta(x>y) \ ,
\end{align} 
 that is a part of ERBL \cite{Efremov:1979qk,Lepage:1980fj} evolution kernel.
   For an analog of TMDA in the $p^2=0$ limit, we have
    \begin{align}
\Psi^{\rm 2-curr} (x, k_\perp;y) =   \frac{g^2}{16\pi^3}  
\frac{V(x,y)} {(k_\perp^2+m^2 )^2 }   \  . 
\end{align} 
A further step is   a  superposition 
model in which the $yp, \bar y p$ states enter with the weight $\varphi_0 (y)$,  a ``primordial'' distribution amplitude.
Then the  model TMDA is given by  a convolution 
    \begin{align}
\Psi^{\rm conv}  (x, k_\perp) = \frac{g^2}{16\pi^3}  \,  \frac1{( k_\perp^2+m^2 )^2 }   
\int_0^1 {V(x,y)}  \, \varphi_0 (y) \, dy  \  .
\label{psi3}
\end{align} 
The integral producing DA in this case converges to give
$\varphi^{\rm conv}  (x) = {g^2}/{(8\pi^2m^2)} \,  \delta \varphi (x)$, where 
   \begin{align}
 \delta \varphi (x) \equiv [V \otimes \varphi_0] \, (x) =  \int_0^1 {V(x,y)}  \, \varphi_0 (y) \, dy  \ , 
\end{align} 
that  has the meaning of  a correction to $\varphi_0 (x)$ generated  by the simplest exchange interaction.
However, the $k_\perp^2$  moment, and all higher $k_\perp^2$ moments  of $\Psi^{\rm conv}  (x, k_\perp)$ 
diverge, which is reflected by  $\ln z_\perp^2$ terms in the 
 expansion of the relevant IDA 
   \begin{align}
\varphi^{\rm conv} (x, z_\perp)  =   m z_\perp  K_1 (mz_\perp) \,  \varphi^{\rm conv}  (x) 
\ , 
\end{align} 
since   $a K_1(a) =1 + a I_1 (a) \ln a $ + analytic terms  for small $a$.

 \subsection{Generating  hard  tail}

\begin{figure}[h]
   \centerline{\includegraphics[width=1.6in]{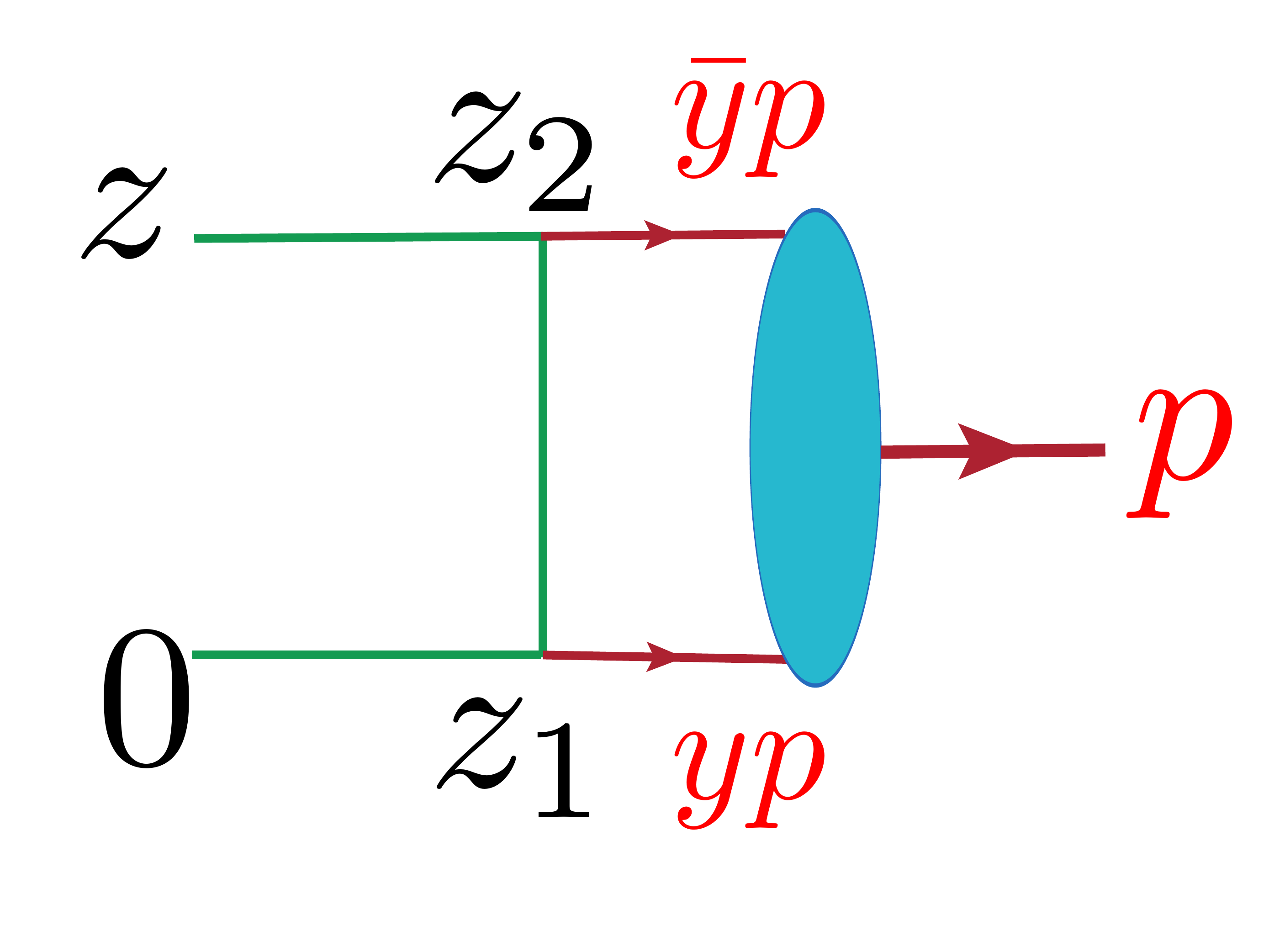}}
   \caption{Hard tail model. 
   \label{bilocal}}
   \end{figure}

 Thus,  an exchange of a ``gluon''   
  has 
converted a superposition of  collinear (to $p$)  ``quark'' states   into a state that 
has  $1/(k_\perp^2+m^2)^2$  dependence on the transverse momentum $k_\perp$.
 We may also assume that the initial fields at $z_1$ and $z_2$ 
are   described by 
some   ``primordial'' bilocal function 
 \mbox{$B_0 (y, (z_1-z_2)^2/4)$}
  corresponding to a soft 
 TMDA $\Psi_0 (y, k_\perp)\equiv \psi_0 (x, k_\perp^2)/\pi$ (see Fig. \ref{bilocal}). 
   To concentrate on virtuality effects induced by $B_0$, 
  we use  $m=0$ and \mbox{$p^2=0$}.   
       Then the  
     generated hard TMDA   is  given by 
      \begin{align}
  \Psi^{B_0}   (x, k_\perp) =   & \frac{  g^2 }{16  \pi^3 k_\perp^2 } \, 
   \int_0^1 dy    \left [  \int_0^1 d \xi  \,        
     \psi_0 \left  (y, \frac{\xi  k_\perp^2}{ V(x,y} \right )   \right ]
     \ . 
     \label{tail}
  \end{align} 
     The term in square brackets may be  written as 
         \begin{align}
  \Biggl [ \cdots \Biggr ]  = \frac{ V(x,y} { k_\perp^2} \left \{
  {\varphi_0 (y)}- \int^\infty_{k_\perp^2/V(x,y) }
     \psi_0 (y, { k'_\perp}^2   )  \, d  {k'_\perp}^2 
    \right \} 
 \ ,
  \end{align} 
where  $\varphi_0 (y)$ is   
the primordial distribution $\Psi_0 (y, k'_\perp)$ integrated
over  all the transverse momentum plane.
 Hence,  for large $k_\perp$, the 
leading $1/k_\perp^4$  term is  determined by 
the DA  $\varphi_0 (y)$ only.
A particular shape  of the \mbox{$k_\perp$-dependence} of  the soft TMDA 
  $\Psi _0(y, k_\perp)$  affects  only 
 the  subleading $\sim [V \otimes \psi_0] (x,k_\perp^2)/k_\perp^2$ term. 
 The form of $k_\perp$ dependence of   $\Psi _0(y, k_\perp)$  is also essential
for the behavior  of  $\Psi^{B_0}  (x, k_\perp)$  
 term at  small $k_\perp$.  In particular,  
 we have 
      \begin{align}
  \Biggl [ \cdots \Biggr ] _{k_\perp=0}  = 
     \psi_0 (y,  k_\perp^2=0   )   
 \ ,
\label {small_k}
  \end{align} 
which gives, e.g., $\varphi_0 (y)/ \Lambda^2$ in the Gaussian model
 (\ref{Gaussian}).   
 
 \subsection{Hard  tail for spin-1/2 quarks}
 
 In case of spin-1/2 quarks interacting $via$ a (pseudo)scalar  gluon field
 (``Yukawa'' gluon model), Eq. (\ref{tail}) is modified by 
 an extra $k_\perp^2$ factor coming from the numerator spinor trace, which  
leads to    $1/(k_\perp^2+m^2)$ dependence in the correction (\ref{psi3})   to 
the TMDA. 
It results in  a $\varphi^{\rm conv}_Y  (x, z_\perp)  \sim     K_0 (mz_\perp) \, \delta \varphi (x)$
term for the  IDA. The  logarithmic divergence for $z_\perp=0$ of this  outcome 
 corresponds to  evolution of the DA.
 In the $B_0$ model, we have  (switching  to $\varphi   (x, z_\perp)   \to \varphi   (x, z_\perp^2)$ in our 
 notations below)    
   \begin{align}
  \varphi^{B_0}_Y   (x, z_\perp^2) =   & \frac{  g^2 }{16  \pi^2  } 
   \int_0^1 dy   \,V(x,y) 
   \int_1^\infty  \frac{d \nu}{\nu} \, \varphi_0 
   \Bigl (y, \nu \, z_\perp^2 \,V(x,y) \Bigr ) 
     \ . 
     \label{tailY}
  \end{align} 
Substituting  formally 
 $ \varphi_0 (y, z_\perp^2)$ by $\varphi_0 (y)$  
 in   the \mbox{$z_\perp^2\to 0$}  limit, 
we get a   logarithmically divergent integral over $\nu$.  
 In fact, for a function $ \varphi_0 (y, z_\perp^2)$ that  rapidly decreases 
 when  \mbox{$z_\perp^2 \gtrsim 1/\Lambda^2$,}  
one gets $\ln (z_\perp^2 \Lambda^2)$ as a factor accompanying 
 the convolution of $V(x,y)$ and $\varphi_0 (y)$. 
 Hence, the pion  size  cut-off contained in the primordial 
 distribution provides the scale in  $\log (z_\perp^2)$,
 and we may keep the hard quark propagators massless.
 This cut-off also results in a finite value of 
 $ \Psi^{B_0}_Y $ in the  formal $k_\perp \to 0$ limit:
   \begin{align}
  \Psi^{B_0}_Y   (x, k_\perp=0) =   & \frac{  g^2 }{16  \pi^2} \, 
   \int_0^1 dy       \, 
     \Psi_0 \left  (y,  k_\perp=0 \right )   
     \ . 
     \label{small_k}
  \end{align} 
  Thus, the  \mbox{$\Psi^{\rm conv}_Y  (x, k_\perp) \sim \delta \varphi (x) 
/k_\perp^2$} singularity of the
``collinear model''
\mbox{$\Psi_0 (y, k_\perp) = \varphi_0 (y) \, \delta (k_\perp^2)/\pi$}  
converts  
into a  constant $1/\Lambda^2  $  in the   Gaussian model. 
Note also that 
 the overall 
 factor in  \mbox{Eq. (\ref{small_k})}  then 
contains  the $x$-independent integral of $\varphi_0 (y)$, i.e. 
$f_\pi$, rather than the  convolution $\delta \varphi (x) $ 
as 
in Eq. (\ref{psi3}).

Concluding, we  emphasize that 
the VDA  approach provides an  unambiguous prescription      of  
  { generating}
hard-tail terms like $\Psi^{B_0}(x, k_\perp)$    from a soft
primordial distribution 
$\Psi_0 (y, k_\perp)$.    
A subject for  future studies is to use this strategy for building hard tail models in 
 case  of  QCD.

 \section{Summary and outlook}
 \label{Summary} 
 
 In the present paper, we  outlined a 
 new approach to transverse momentum dependence 
 in hard processes.  Its starting point, just like in the OPE formalism,
  is the use 
 of coordinate representation. At handbag level,  the structure  
 of a hadron with momentum $p$  is described  
 by   a matrix element 
 of the bilocal operator ${\cal O} (0,z)$,   
 treated as a function of $(pz)$ and $z^2$.  It is parametrized 
 through a    {\it virtuality distribution}  $\Phi (x, \sigma)$,  in which 
 the variable $x$ is Fourier-conjugate to $(pz)$,  and has the usual 
 meaning of a  parton  momentum fraction.  Another parameter, $\sigma$,  is 
conjugate to $z^2$ through an analog  of   Laplace transform.
 
 Projecting ${\cal O} (0,z)$ onto  a spacelike interval with   $z^+=0$,  
 we   introduce {\it transverse momentum distributions} 
 $\Psi (x, k_\perp)$  and show that they can be written  
 in terms of  virtuality distributions $\Phi (x, \sigma)$. 
This fact opens  the possibility 
 to  convert the results of covariant calculations,  
written in terms  of $\Phi (x, \sigma)$,  into expressions 
involving  $\Psi (x, k_\perp)$. 
This procedure  being a crucial feature of our approach,   is illustrated  
 in the present paper  
by its application to  hard exclusive transition process
 $\gamma^* \gamma \to \pi^0$  at the handbag level
 (which is analogous to the 2-body Fock state approximation).
   Starting with scalar toy models, we then extend the analysis  
   onto the case of spin-1/2 quarks  and vector  gluons.

We propose a few simple models for soft VDAs/TMDAs,
and  use them  for comparison of VDA results with experimental 
(BaBar and BELLE)  data 
on  the pion transition form factor. 
 
A  natural  next step
is going beyond the handbag approximation. 
 In QCD, an important feature is that  quark-gluon interactions 
generate a hard $\sim 1/k_\perp^2$ tail for TMDAs. 
To demonstrate the capabilities  of the VDA approach 
in this direction, 
we  describe 
the basic elements of generating hard tails from soft primordial 
TMDAs.

Another direction for  future studies is an extension of the 
VDA approach onto inclusive reactions, such as 
 Drell-Yan and SIDIS processes. 
 In particular, we envisage building  VDA-based models 
 for soft parts of TMDs that would have a non-Gaussian
 behavior at large $k_\perp$
 (the need for such models was recently emphasized
 by several authors 
 \cite{Collins:2013zsa, Sivers:2013eca,Schweitzer:2012hh}). The VDA approach would also allow to 
  self-consistently
 generate hard tails from these soft TMDs.


\section*{Acknowledgements}

I thank I. Balitsky, G. A.  Miller,  A.H. Mueller, 
A. Prokudin, A. Tarasov  and  C. Weiss for discussions.
This work is supported by Jefferson Science Associates,
 LLC under  U.S. DOE Contract \#DE-AC05-06OR23177
 and by U.S. DOE Grant \#DE-FG02-97ER41028. 





\begin{thebibliography}{10}
\bibitem{Mulders:1995dh} 
  P.~J.~Mulders and R.~D.~Tangerman,
  Nucl.\ Phys.\ B {\bf 461}, 197 (1996)



\bibitem{Georgi:1976ve} 
  H.~Georgi and H.~D.~Politzer,
  Phys.\ Rev.\ D {\bf 14}, 1829 (1976).



\bibitem{Efremov:1976ih} 
  A.~V.~Efremov and A.~V.~Radyushkin,
  Lett.\ Nuovo Cim.\  {\bf 19}, 83 (1977).



\bibitem{Lepage:1980fj} 
  G.~P.~Lepage and S.~J.~Brodsky,
  Phys.\ Rev.\ D {\bf 22}, 2157 (1980).



\bibitem{delAguila:1981nk} 
  F.~del Aguila and M.~K.~Chase,
  Nucl.\ Phys.\ B {\bf 193}, 517 (1981).



\bibitem{Braaten:1982yp} 
  E.~Braaten,
  Phys.\ Rev.\ D {\bf 28}, 524 (1983).



\bibitem{Kadantseva:1985kb} 
  E.~P.~Kadantseva, S.~V.~Mikhailov and A.~V.~Radyushkin,
 Yad.\ Fiz.\  {\bf 44}, 507 (1986)
 [ Sov.\ J.\ Nucl.\ Phys.\  {\bf 44}, 326 (1986)].



\bibitem{Musatov:1997pu} 
  I.~V.~Musatov and A.~V.~Radyushkin,
  Phys.\ Rev.\ D {\bf 56}, 2713 (1997)



\bibitem{Radyushkin:1977gp} 
  A.~V.~Radyushkin,
  {\em JINR report P2-10717 (unpublished);{\rm hep-ph/0410276}
  (English translation)}   (1977)



\bibitem{Efremov:1979qk} 
  A.~V.~Efremov and A.~V.~Radyushkin,
  Phys.\ Lett.\ B {\bf 94}, 245 (1980).



\bibitem{Chernyak:1977fk} 
  V.~L.~Chernyak, A.~R.~Zhitnitsky and V.~G.~Serbo,
  JETP Lett.\  {\bf 26}, 594 (1977)
  [Pisma Zh.\ Eksp.\ Teor.\ Fiz.\  {\bf 26}, 760 (1977)].



\bibitem{Lepage:1979zb} 
  G.~P.~Lepage and S.~J.~Brodsky,
  Phys.\ Lett.\ B {\bf 87}, 359 (1979).



\bibitem{Sudakov:1954sw} 
  V.~V.~Sudakov,
  Sov.\ Phys.\ JETP {\bf 3}, 65 (1956)



\bibitem{Nachtmann:1973mr} 
  O.~Nachtmann,
  Nucl.\ Phys.\ B {\bf 63}, 237 (1973).



\bibitem{Radyushkin:1983wh} 
  A.~V.~Radyushkin,
  Phys.\ Lett.\ B {\bf 131}, 179 (1983).



\bibitem{Radyushkin:1983ea} 
  A.~V.~Radyushkin,
  Theor.\ Math.\ Phys.\  {\bf 61}, 1144 (1985)



\bibitem{Radyushkin:1997ki} 
  A.~V.~Radyushkin,
  Phys.\ Rev.\ D {\bf 56}, 5524 (1997)



\bibitem{Novikov:1983jt} 
  V.~A.~Novikov, M.~A.~Shifman, A.~I.~Vainshtein, M.~B.~Voloshin and V.~I.~Zakharov,
  Nucl.\ Phys.\ B {\bf 237}, 525 (1984).



\bibitem{Efremov:1978fi} 
  A.~V.~Efremov and A.~V.~Radyushkin,
  JINR-E2-11535 (1978).



\bibitem{Efremov:1978xm} 
  A.~V.~Efremov and A.~V.~Radyushkin,
  Theor.\ Math.\ Phys.\  {\bf 44}, 774 (1981)



\bibitem{Efremov:1980ub} 
  A.~V.~Efremov and A.~V.~Radyushkin,
  Riv.\ Nuovo Cim.\  {\bf 3N2}, 1 (1980).


\bibitem{Fock:1937aa}
V.~Fock, {\em Sowjet. Phys.} {\bf 12}, p. 404  (1937).


\bibitem{PhysRev.82.664} 
  J.~S.~Schwinger,
  Phys.\ Rev.\  {\bf 82}, 664 (1951).



\bibitem{Aubert:2009mc} 
  B.~Aubert {\it et al.}  [BaBar Collaboration],
  Phys.\ Rev.\ D {\bf 80}, 052002 (2009)



\bibitem{Uehara:2012ag} 
  S.~Uehara {\it et al.}  [Belle Collaboration],
  Phys.\ Rev.\ D {\bf 86}, 092007 (2012)



\bibitem{Radyushkin:2009zg} 
  A.~V.~Radyushkin,
  Phys.\ Rev.\ D {\bf 80}, 094009 (2009)



\bibitem{Polyakov:2009je} 
  M.~V.~Polyakov,
  JETP Lett.\  {\bf 90}, 228 (2009)



\bibitem{Li:1992nu} 
  H.~-n.~Li and G.~F.~Sterman,
  Nucl.\ Phys.\ B {\bf 381}, 129 (1992).



\bibitem{Collins:2013zsa} 
  J.~Collins,
  Int.\ J.\ Mod.\ Phys.\ Conf.\ Ser.\  {\bf 25}, 1460001 (2014)



\bibitem{Sivers:2013eca} 
  D.~Sivers,
  Int.\ J.\ Mod.\ Phys.\ Conf.\ Ser.\  {\bf 25}, 1460002 (2014)



\bibitem{Schweitzer:2012hh} 
  P.~Schweitzer, M.~Strikman and C.~Weiss,
  JHEP {\bf 1301}, 163 (2013)

.

\end{thebibliography}

 \end{document}